\documentclass[amsmath,aps,prd,twocolumn,superscriptaddress,showpacs,showkeys]{revtex4}

\newcommand{\Bsymmass}{M_{0}}
\newcommand{\Nmag}{\mu_{N}}
\newcommand{\Bmag}{\mu_{0}}
\newcommand{\bracket}[3]{\langle #1 | #2 | #3 \rangle}

\begin{document}

\title{Strange magnetic moment of the nucleon and SU(3) breaking: \\ group theoretical approach\footnote{Work supported in part by the Alexander von Humboldt Foundation and by BMBF.}}
\author{D. Jido}
\email{jido@ph.tum.de}
\author{W. Weise}
\affiliation{Physik-Department, Technische Universit\"at M\"unchen, D-85747 Garching, Germany}
\date{\today }

\begin{abstract}
An extended group-theoretical approach to magnetic moments of the octet baryons is proposed with the aim of extracting the strange magnetic moment of the nucleon. Special attention is given to flavor SU(3) breaking. In this approach, isoscalar and isovector magnetic moments are treated separately in view of their different behavior under SU(3) breaking. We conclude that the anomalous magnetic moment associated with the flavor singlet current is small. Together with the small isoscalar anomalous magnetic moment of the nucleon, this implies suppression of the strange magnetic moment of the proton which is found to be small and positive,
$\mu^{(s)} = (0.16 \pm 0.03)\,\mu_N$ in units of the nuclear magneton. 
\end{abstract}
\maketitle

\section{Introduction}

The strange quark content of the nucleon magnetic moment is an important issue in understanding the structure of the nucleon. 
Recent investigations of the strange quark contribution to the electromagnetic proton form factors have been performed with parity violating elastic electron-proton scattering at JLab (HAPPEX) 
\cite{Aniol:2000at,Aniol:2005zg}, MIT-Bates (SAMPLE) \cite{Spayde:2003nr,Ito:2003mr} and 
Mainz (A4) \cite{Maas:2004ta,Maas:2004dh}. 
For example, a value extracted from the SAMPLE experiment is $G^{s}_{M}(Q^{2}=0.1\,GeV^2) =0.37\pm 0.20 \pm 0.26 \pm 0.07$ in units of the nuclear magneton, $\Nmag\equiv e\hbar/(2M_{p}c)$. 
A combined fit of all the measurements at $Q^{2} \sim 0.1$ GeV$^{2}$ suggests positive values  $G^{s}_{M}=+0.55\pm0.28$, still leaving open the possibility of $G^{s}_{M}=0$ at 95\% confidence level~\cite{Aniol:2005zg}. 
There have been many theoretical attempts to calculate the strange magnetic moment of nucleon,
for instance, in a vector meson dominance picture~\cite{Jaffe:1989mj},  
Skyrme soliton models~\cite{Park:1990is,Park:1991fb}, 
a dispersion relation analysis~\cite{Hammer:1995de,Forkel:1995ff}, 
meson-exchange model~\cite{Meissner:1997qt},
chiral perturbation theory~\cite{Hemmert:1998pi}, chiral quark soliton model \cite{Kim:1998gt},
QCD equalities~\cite{Leinweber:1995ie} and lattice computation~\cite{Leinweber:2004tc}.

This paper presents an investigation of the limits set for the strange magnetic moment of the nucleon within the framework of a group theoretical approach. These considerations do not rely on specific assumptions about the detailed intrinsic spin structure of the nucleon. The constraints imposed by the SU(3) symmetry breaking pattern and possible mixing of the octet with 
singlet, decuplet, anti-decuplet and 27-plet
representations in the magnetic moment operators will be examined in a framework based on group theoretical relations, without explicit reference to quark model wave functions. 

The group theoretical approach for the baryon magnetic moments was first developed in Ref.\cite{Coleman:1961jn} for the SU(3) symmetric limit, leading to the Coleman-Glashow relations. In Ref.\cite{Caldi:1974ta} corrections from chiral perturbation theory were included. Effects from mixing of representations other than the octet in the magnetic moment were discussed using the spectrum-generating SU(3) approach in 
Ref.\cite{Bohm:1977na}. 
The SU(6) quark model is also along this line. 

The important point is that, under the SU(3) breaking, we should consider not only the mixing of the singlet operator but also the separate treatment of the isovector and isoscalar magnetic moments, since the SU(3) breaking contributes differently to each isospin sector.
Throughout this argument we assume isospin symmetry. 

\section{SU(3) framework}
\subsection{Basic operators}
Let us consider the flavor singlet vector current $J_{[1]}^{\mu}$ together with the octet current $J_{[8]}^{a,\mu}$ $(a=1,\dots,8)$. Each component of the vector currents is written in terms of the quark fields as
\begin{equation}
  J_{[8]}^{a,\mu} = \bar q \gamma^{\mu} \frac{\lambda^{a}}{2} q ,
  \hspace{0.7cm}
  J_{[1]}^{\mu} = \bar q \gamma^{\mu} \frac{I}{3} q,
  \label{eq:U3curr}
\end{equation}
where $\lambda^{a}$ are the Gell-Mann matrices normalized as ${\rm Tr}[\lambda^{a}\lambda^{b}] = 2 \delta^{ab}$ and $ I$ denotes the $3 \times 3$ identity matrix.

The electromagnetic current is given by the following linear combination of the 3rd and 8th components of the octet current: 
\begin{equation}
  J^{\mu}_{\gamma} = 
  J^{3,\mu}_{[8]} + \frac{1}{\sqrt 3} J^{8,\mu}_{[8]},
  \label{eq:elemagcurrent}
\end{equation}
in which the first and second term includes the isovector and isoscalar current, respectively. 
The pure strange vector current is written as a linear combination of the flavor octet and singlet currents:
\begin{equation}
  J^{\mu}_{(s)} = 
  J^{\mu}_{[1]} - \frac{2}{\sqrt 3} J^{8,\mu}_{[8]}.
  \label{eq:strangecurrent}
\end{equation}

The Dirac couplings of the baryons to the vector current~(\ref{eq:U3curr}) are fixed by the extended gauge coupling based on charge and flavor conservation:
\begin{equation}
  {\cal L}_{\rm D} = 
  ie\left( {\rm Tr}\left[\bar B \gamma_{\mu} [ \lambda^{a},B] \right] J_{[8]}^{a,\mu} + 
  {\rm Tr}\left[\bar B \gamma_{\mu}B\right] J_{[1]}^{\mu}\right),
  \label{eq:elecoup}
\end{equation}
where the summation over $a$ is taken from 1 to 8, and
the baryon fields are expressed in $3\times 3$ matrix form as
\begin{equation}
   B = 
   \left(
   \begin{array}{ccc}
       \frac{1}{\sqrt{2}} \Sigma^0 + \frac{1}{\sqrt{6}} \Lambda &
       \Sigma^+ & p \\
       \Sigma^- & -\frac{1}{\sqrt{2}} \Sigma^0 + 
       \frac{1}{\sqrt{6}} \Lambda  & n\\
       \Xi^- & \Xi^0 &- \frac{2}{\sqrt{6}} \Lambda 
   \end{array}
   \right).  \label{eq:barymat}
\end{equation}
The octet and singlet currents in Eq.(\ref{eq:elecoup}) couple equally to the baryons. The ratio of the isoscalar and isovector couplings is fixed by charge and flavor conservation. The OZI rule is automatically satisfied in this form.
On the other hand, the tensor (Pauli) coupling in which the anomalous magnetic moments appear, is manifestly gauge invariant, so that there are no such restriction among the flavor octet and singlet currents.

Consider now the magnetic moment operators for the octet baryons in the SU(3) basis:  $\hat \mu^{a}_{[8]}$ $(a=1,\dots,8)$ for the octet and $\hat\mu_{[1]}$ for the singlet. 
SU(3) symmetry implies that the baryon matrix elements of each component of the magnetic moment operator are parametrized by three independent quantities $F$, $D$ and $S$:
\begin{eqnarray}
  \bracket{B}{\hat \mu^{a}_{[8]}}{B} &=& \left\{ F\, {\rm Tr}\! \left[\bar B [\lambda^{a}, B]\right] \nonumber \right. \\ &&
  \left.+D\, {\rm Tr}\! \left[\bar B \{\lambda^{a}, B\} \right] \right\} \Bmag 
   \label{eq:su3magoct} \\ 
\bracket{B}{\hat \mu_{[1]}}{B} &=&
  S\, {\rm Tr}\! \left[\bar B B\right] \, \Bmag, 
  \label{eq:su3magsing}
\end{eqnarray}
where the magnetic moments are measured in units of the SU(3) symmetric magneton defined by $\Bmag \equiv e\hbar/(2 \Bsymmass c)$ with the octet baryon mass in the exact SU(3) limit,  $\Bsymmass=1152$ MeV. The derivation of the SU(3) symmetric baryon mass is given in Appendix.   
We assume that the ground state baryons are in the octet representation of the SU(3) group without mixing of the other representations, since the observed octet baryon masses satisfy very well the Gell-Mann -  Okubo relation, obtained by introducing SU(3) breaking on the mass operator.

The strange magnetic moment of the proton is given in terms of the parameters $S$, $F$ and $D$ as
\begin{equation}
\langle p| \hat \mu^{(s)} | p \rangle = \left[ S - 2 \left(F-\frac{D}{3}\right) \right] \Bmag \ , \label{eq:strangeMM}
\end{equation}
where the operator of the strange magnetic moment is defined in the SU(3) basis as
\begin{equation}
  \hat \mu^{(s)} =  \hat\mu_{[1]} - \frac{2}{\sqrt{3} }\hat \mu^{8}_{[8]}  ,
  \label{eq:stmagope}
\end{equation}
analogous to the strange quark current (\ref{eq:strangecurrent}).
Hereafter we shall determine the parameters in Eqs.(\ref{eq:su3magoct}) and (\ref{eq:su3magsing}) from the observed magnetic moments of the octet baryons. 

In the exact SU(3) limit, the operator of the magnetic moment associated with the electromagnetic current (\ref{eq:elemagcurrent}) is given by 
\begin{equation}
  \hat \mu^{\gamma}_{\rm SU(3)} = 
     \hat\mu^{3}_{[8]}  + \frac{1}{\sqrt 3} \hat\mu^{8}_{[8]} . 
     \label{eq:su3magope}
\end{equation}    
Since the electromagnetic current does not have a flavor singlet component, the magnetic moments of the octet baryons are all expressed by the two parameters $F$ and $D$.  This leads to the Coleman-Glashow relations \cite{Coleman:1961jn}:
\begin{equation}
  \begin{array}{ccc}
     \mu_{\Sigma^{+}}=\mu_{p},\ \ & 
     \mu_{\Lambda} = {\textstyle\frac{1}{2}} \mu_{n},\ \  &
     \mu_{\Xi^{0}} = \mu_{n},\ \  \\
     \mu_{\Sigma^{0}}=-{\textstyle\frac{1}{2}}\mu_{n},\ \  &
     \multicolumn{2}{c}{\mu_{\Xi^{-}}=\mu_{\Sigma^{-}}=-(\mu_{p}+\mu_{n})}.\ \ 
  \end{array}
\end{equation}
The observed magnetic moments satisfy these relations  approximately at the level of 20\,\%.
\subsection{Symmetry breaking scenarios}
The deviations from the Coleman-Glashow relations are presumably caused by flavor SU(3) breaking. Improvements can be introduced by taking into account the following sources of SU(3) symmetry breaking. First of all, the corresponding ratio of the isovector and isoscalar parts in the magnetic moment operators can deviate from the ratio in the electromagnetic current since the Pauli couplings are renormalized by quark loops in a different way for the isovector and isoscalar parts under the SU(3) breaking \cite{Leinweber:1995ie}. This means that SU(3) relations are satisfied independently in the isovector and isoscalar sectors. Separate SU(3) relations within the octet scheme of the magnetic moment operator can therefore be written as follows:
\begin{equation}
  \mu_{N}^{\rm IV}+\mu_{\Xi}^{\rm IV} = \mu_{\Sigma}^{\rm IV} \ ,
    \hspace{1cm}  
  \mu_{N}^{\rm IV}-\mu_{\Xi}^{\rm IV} = \sqrt 3\, \mu_{\Sigma^{0}\Lambda}
  \label{eq:relationIV}
\end{equation}
for the isovector (IV) and
\begin{equation}
 \mu_{N}^{\rm IS} + \mu_{\Xi}^{\rm IS} =
  \mu_{\Lambda} =- \mu_{\Sigma}^{\rm IS} 
  \label{eq:relationIS}
\end{equation}
for the isoscalar (IS) moments. Here the isoscalar $\Sigma$ magnetic moment is defined as $\mu_{\Sigma}^{\rm IS}=\frac{1}{2}(\mu_{\Sigma^{+}}+\mu_{\Sigma^{-}})$. 
The first relation of isovector (\ref{eq:relationIV}) and the isoscalar relation (\ref{eq:relationIS}) were also obtained in Refs.\cite{Franklin:1969sd} and \cite{Caldi:1974ta}, respectively.
The relations (\ref{eq:relationIV}, \ref{eq:relationIS}) are satisfied at the 10\% accuracy level. Hereafter, we perform fits separately in the isovector and isoscalar magnetic moments.

Another source of the SU(3) breaking is mixing with the magnetic moment operators which belong to irreducible representations of SU(3) other than the octet. 
All possible representations mixing with the operator for the octet baryon are singlet, decuplet, anti-decuplet and 27-plet, since decomposition of the direct product of two octets to a direct sum is given by $\bf 8 \otimes 8 = 1 \oplus 8_s \oplus 8_a \oplus 10 \oplus \overline{10} \oplus 27$.
This kind of mixing results, for example, when considering different meson masses in meson loops. 

With inclusion of SU(3) breaking effects the magnetic moment operators can in general be written as:
\begin{eqnarray}
\hat\mu_{\rm SB}^{\gamma} &= &
  v_{8}\, \hat\mu^{3}_{[8]} + 
  \frac{s_{8}}{\sqrt{3}}\,  \hat\mu^{8}_{[8]}+ 
  s_{1}\, \hat \mu_{[1]} \nonumber \\
  && + v_{10} (\hat\mu_{[10]} + \hat\mu_{[\overline{10}]}) + 
  v_{27}\, \hat\mu_{[27]}^{\rm IV}+
  \frac{s_{27}}{\sqrt 3} \, \hat\mu_{[27]}^{\rm IS},
  \label{eq:magope}
\end{eqnarray}
where $\hat\mu_{[10]}$, $\hat\mu_{[\overline{10}]}$ and $\hat \mu_{[27]}$ is the magnetic moment operators belonging to the decuplet, anti-decuplet and 27-plet representations, respectively. 
In Eq.(\ref{eq:magope}), we have already taken into account the hermitian nature of the magnetic moment operator. The 27-plet operator with $I=2$ cannot be induced under isospin symmetry. 
The coefficients $s_{i}$ (isoscalar) and $v_{i}$ (isovector) are dependent on the SU(3) breaking pattern. In the SU(3) symmetric limit, the coefficients satisfy $s_{8}=v_{8}=1$ and $s_{1}=s_{27}=v_{10}=v_{27}=0$ to recover Eq.(\ref{eq:su3magope}).

Evaluating the baryon matrix elements of the magnetic moment operator (\ref{eq:magope}), we find the following expressions for the magnetic moments in terms of the group theoretical parameters:
\begin{eqnarray}
  \mu_{N}^{\rm IV} &=& v_{8}(F+D)+v_{10} K + v_{27} L,\  \\
  \mu_{\Xi}^{\rm IV}&=& v_{8}(F-D)+v_{10} K -v_{27} L,\\
  \mu^{\rm IV}_{\Sigma}&=& v_{8}\, 2F - v_{10} K,  \\
 \sqrt 3 \mu_{\Sigma^{0}\Lambda} &=& v_{8}\, 2D - v_{27} \, 3L,
\end{eqnarray}
for isovector and
\begin{eqnarray}
\mu_{N}^{\rm IS} &=& s_{8} ( F-{\textstyle \frac{1}{3}} D)+ s_{1} S^{(a)}+ s_{27}\,  {\textstyle \frac{1}{2}} L, \\
\mu_{\Lambda} &=& -s_{8}\, {\textstyle \frac{2}{3}}D +s_{1} S^{(a)} - s_{27}\, {\textstyle \frac{3}{2}}  L\\
\mu^{\rm IS}_{\Sigma}&=& s_{8}\, {\textstyle \frac{2}{3}}D +s_{1} S^{(a)} - s_{27}\, {\textstyle \frac{1}{6} } L, \\
\mu_{\Xi}^{\rm IS} &=&s_{8}( -F-{\textstyle \frac{1}{3}}D)+ s_{1} S^{(a)}+s_{27}\, {\textstyle \frac{1}{2}}  L,
\end{eqnarray}
for isoscalar. The isoscalar $\Sigma$ magnetic moment is again given as $\mu_{\Sigma}^{\rm IS}=\frac{1}{2}(\mu_{\Sigma^{+}}+\mu_{\Sigma^{-}})$, which is the case when the $I=2$ component from the 27-plet operator is negligible.
The parameters, $F$, $D$, are defined in Eq.(\ref{eq:su3magoct}) as reduced matrix elements of the antisymmetric and symmetric octet operators, respectively. The reduced matrix elements for
the singlet, decuplet, anti-decuplet and 27-plet, $S^{(a)}$, $K$ and $L$,
contribute only to the anomalous moments since the electromagnetic current has no component of 
these representations. 
The normal part of the singlet operator is fixed by the Dirac coupling (\ref{eq:elecoup}) as one unit of $\Bmag$. Therefore the singlet parameter $S$ defined in Eq.(\ref{eq:su3magsing}) is given as $S=1+S^{(a)}$. 
The reduced matrix elements of the $\hat\mu_{[10]}$, $\hat\mu_{[\overline{10}]}$ and $\hat\mu_{[27]}$ operators are normalized as $\langle p |\hat\mu_{[10]}| p \rangle = - \langle p |\hat\mu_{[\overline{10}]}| p \rangle = \frac{i}{2} K$ and $\langle p | \hat\mu_{[27]}^{\rm IV}|p \rangle = L$, respectively. The matrix elements for other states are obtained with the SU(3) Clebsch-Gordan coefficients given in Ref.\cite{McNamee:1964aa}.

\begin{table}
\caption{Fitted parameters in units of $\Bmag$. \label{tab:param}}
\begin{ruledtabular}
\begin{tabular}{ccccccccc}
   &\multicolumn{4}{c}{isovector}& \multicolumn{4}{c}{isoscalar}\\
    \cline{2-5}\cline{6-9}
   &  $v_{8}F$  & $v_{8}D$ & $v_{10} K $ & $v_{27} L$  & 
      $s_{8}F$  & $s_{8}D$ & $s_{1} S^{(a)}$ &$s_{27} L$ \\
\hline
Fit 1  &1.160 & 1.661  &---&---& 0.853 & 1.117 &  --- &---\\
Fit 2  &1.160 & 1.661  &---&---& 0.853 & 1.144 &  0.045 &---\\
Fit 3  &1.160 & 1.661  &0.100&-0.033 & 0.853 & 1.130& 0.048 & 0.031 \\
\end{tabular}
\end{ruledtabular}
\end{table}

We now perform separate fits of the isovector and isoscalar magnetic moment 
in the following three cases: Fit 1) with only the octet operator, setting 
$s_{1}=v_{10}=s_{27}=v_{27}=0$ 
; Fit 2) with the octet and singlet operators, setting 
$v_{10}=s_{27}=v_{27}=0$ 
; Fit 3) with the complete set of operators. 
For the cases of Fit1 and Fit2 we use the least square method. 
For Fit 3, the equations are completely solved, given equal numbers of input and output quantities.    
%
The fitted parameters in each procedure are listed in Table \ref{tab:param}, and
the values of the magnetic moments reproduced by
these fits are summarized in Table \ref{tab:magiso}.

\begin{table}
\caption{Magnetic moments in the isospin basis in units of $\Nmag$.
Experimental data translated to the isospin basis (exp.) are obtained from the magnetic moments in the particle basis listed in Ref.\cite{Eidelman:2004wy}.   Parenthesis denote uncertainties estimated from the upper limits of the experimental errors given in Ref.\cite{Eidelman:2004wy}.
\label{tab:magiso}}
\begin{ruledtabular}
\begin{tabular}{ccccc}
 &\multicolumn{4}{c}{isovector}\\
 \cline{2-5}
 & $N$ &$\Sigma$ &$\Xi$ &$\Sigma^{0}$-$\Lambda$  \\
 \hline
  exp. & $2.35294$ &$1.809(18)$&$-0.2997(83)$&$1.61(8)$\\
  \hline
  Fit 1, 2 & $2.299$ & $1.890$ &$ -0.408$ & $1.563$ \\
  Fit 3 & $2.353$ & $1.809$ &$ -0.300$ & $1.610$ \\
  \hline\hline
&\multicolumn{4}{c}{isoscalar}\\
 \cline{2-5}
 & $N$   &$\Lambda$&$\Sigma$ &$\Xi$  \\
 \hline
  exp. & $0.43990$&$-0.613(4)$ &$0.649(18)$ &$-0.9504(83)$\\
  \hline
  Fit 1 & $0.392$ & $-0.607$ & $0.607$ & $-0.999$\\
  Fit 2 & $0.421$ & $-0.585$ & $0.658$ & $-0.969$\\
  Fit 3 & $0.440$ & $-0.613$ & $0.649$ & $-0.950$
\end{tabular}
\end{ruledtabular}
\end{table}

The results of our fits indicate that the contributions from the 
$\bf 1$, $\bf 10$, $\bf \overline{10}$ and $\bf 27$ representations
%
are small in comparison with the overall magnitudes of the magnetic moments. On the other hand,  the deviation of the ratio of $v_{8}$ and $s_{8}$ from 1, also caused by SU(3) breaking, is large.  The ratios obtained by the fits are around 1.4, showing a sizable deviation from unity.  

So far, we have decomposed the observed magnetic moments into the SU(3) irreducible representations.  This procedure is independent of models, assuming octet dominance for the baryon states. 
Next
we estimate the magnitude of the SU(3) breaking effects among the $v_{i}$ and $s_{i}$ coefficients in the operator (\ref{eq:magope}).
For this purpose, we introduce 
a constituent quark 
picture
in which the baryon magnetic moment operator is expressed in terms of the 
normal (Dirac) magnetic moments of the constituents. 
The anomalous moments of constituent quarks are estimated to be negligible up to next-to-leading order of the large $N_{c}$ expansion when taking for guidance the Gerasimov-Drell-Hearn (GDH) sum rule evaluated with a chiral quark model~\cite{Dicus:1992rw}.

The quark magnetic moment is written in matrix form as 
\begin{equation}
  \mu_{q} = \frac{Q_{q}}{2 m_{q}}
\end{equation}
with the quark charge operator
\begin{equation}
   Q_{q} = {e\over 2}\left( \lambda^{3}  + 
       \frac{1}{\sqrt 3} \lambda^{8}\right)
\end{equation}
and the quark masses $m_{q}$ with $q=u,d,s$. Flavor SU(3) breaking is introduced by $m_{s} > m_{u,d}$. The quark masses are expressed with a SU(3) breaking parameter $a$ as
\begin{equation}
   m_{q} = \bar m (1 + a \sqrt 3\, \lambda^{8})  , \label{eq:quarkmass}
\end{equation}
where $\bar m$ denotes an average quark mass $\bar m$.
Keeping the first order in the SU(3) breaking parameter $a$, we obtain
\begin{equation}
  \mu_{q} \simeq  \bar\mu_{q} \left[
  (1-a) \frac{\lambda^{3}}{2} + 
  \frac{1}{\sqrt 3}(1+a) \frac{\lambda^{8}}{2} - a \frac{I}{3}
  \right] , \label{eq:quarkmagsymb}
\end{equation}
where $\bar \mu_{q}$ denotes the SU(3) symmetric quark magneton $e\hbar/(2\bar m c)$. We have used $\lambda^{3}\lambda^{8}=\lambda^{3}/\sqrt 3$ and $\lambda^{8}\lambda^{8}=(2/3)I - \lambda^{8}/\sqrt 3$.
It is important to note that, in Eq.(\ref{eq:quarkmagsymb}), the SU(3) breaking contributes differently to isovector and isoscalar magnetic moments, and also that the SU(3) breaking induces the singlet component. 
Comparing Eqs.(\ref{eq:magope}) and (\ref{eq:quarkmagsymb}), we relate the coefficients of the magnetic moment operator to the SU(3) breaking parameter $a$ as  
\begin{equation}
  \begin{array}{cc}
 v_{8} = 1-a, \   &  s_{8} = 1+a,\  \\
s_{1}=-a, \ & v_{10}=v_{27}=s_{27}=0. \label{eq:assump}
 \end{array}
\end{equation}

Encouraged by the success of the separate fitting procedure, we evaluate the reduced matrix elements $F$, $D$, $S$ in Eqs.(\ref{eq:su3magoct}) and (\ref{eq:su3magsing}) together with the SU(3) breaking parameter $a$ from the fit values given in Table \ref{tab:param}, using the model just explained.
The resulting SU(3) parameters are summarized in Table \ref{tab:su3param}. In this table, $a_{f}$ and $a_{d}$ denote the SU(3) breaking parameters $a$ related to the $F$ and $D$ terms, respectively, and the final estimate for $a$ is given as the average of $a_{f}$ and $a_{d}$. The value of $S$ shown in the table includes additively the normal magnetic moment for the singlet current, which is unity. 
An error of $S$ is estimated by using explicitly $a_{f}$ and $a_{d}$, instead of their average, 
to compute $S$. 

We obtain the SU(3) breaking parameter $a \simeq -0.17$ which corresponds to $m_{u,d}/m_{s} = 0.62$ in Eq.(\ref{eq:quarkmass}). Assuming a light constituent quark mass $m_{u,d}=320$ MeV, the strange quark mass becomes $m_{s}=516$ MeV, consistent with the SU(6) quark model analysis of the baryon magnetic moments~\cite{Teese:1979fk}. 
We also show in Table \ref{tab:magpart} the magnetic moments in the particle basis, evaluated using Eq.(\ref{eq:assump}), with the parameters $F$, $D$, $S$ and $a$ obtained in Fit 2.

\begin{table}
\caption{Reduced matrix elements, $F$, $D$ and $S$, and SU(3) breaking parameter $a$ obtained by the fits and using Eq.(\ref{eq:assump}). Here $S$ includes the normal magnetic moment. The SU(3) breaking parameters $a_{f}$ and $a_{d}$ are obtained by the $F$ and $D$ terms, respectively. The values of $F$, $D$ and $S$ are presented in units of $\Bmag$. The parenthesis denote the estimated errors for the last digits.
 \label{tab:su3param}}
\begin{ruledtabular}
\begin{tabular}{ccccccc}
       &  $F$ & $D$& $S$&$a_{f}$& $a_{d}$ & $a$\\
\hline
Fit 1  &$1.007$& $1.389$ & $1$ &$ -0.152$ & $-0.196$ & $-0.174$\\
Fit 2  &$1.007$& $1.403$ & $1.268(28)$ & $-0.152$ & $-0.184$&$-0.168$\\
Fit 3  &$1.007$& $1.396$ & $1.280(35)$ & $-0.152$ & $-0.190$ & $-0.171$\\
\end{tabular}
\end{ruledtabular}
\end{table}

The flavor singlet anomalous magnetic moment, $S^{(a)}=S-1$, obtained after removing the SU(3) breaking effect, is around $0.27 \,\Bmag$, less than a third of the normal moment,
in contrast to the anomalous part of $F$ in the octet, which is comparable with the normal moment, $F^{(n)}=1/2$. 

\begin{table}
\caption{Magnetic moments of the octet baryons evaluated using Eq.(\ref{eq:assump}) in units of $\Nmag$. The experimental data are taken from Ref.\cite{Eidelman:2004wy}. The parenthesis denote the experimental errors for the last digits.
\label{tab:magpart}}
\begin{ruledtabular}
\begin{tabular}{ccccc}
  & $p$& $n$ &$\Lambda$ &$\Sigma^{0}$-$\Lambda$ \\
\hline
 exp &$2.7928$&$-1.9130   $ &$-0.613(4) $  & $1.61(8)$  \\
\hline
Fit 2 &$2.696$& $ -1.892$& $-0.597$& $1.541$\\
\hline
\hline
  &$\Sigma^{+}$& $\Sigma^{-}$  & $\Xi^{0}$&$\Xi^{-}$\\
\hline
 exp &$2.458(10)$  &$-1.160(25)$  &$-1.250(14$)&$-0.6507(25)$   \\
\hline
Fit 2 &$2.587$& $-1.246$& $-1.339$ & $-0.585$\\
\end{tabular}
\end{ruledtabular}
\end{table}

\subsection{Evaluating the strangeness magnetic moment}

Let us now evaluate the strange magnetic moment of the nucleon 
using Eq.(\ref{eq:strangeMM})
with $F$, $D$ and $S$ obtained by the fits. 
The values obtained by each fit are summarized in Table \ref{tab:strangmag} 
with estimated theoretical errors coming from the evaluation of $S$.
%
%
We find, in Fit 2 and 3, that the strange magnetic moment is positive and small, but still 2.5 times larger in magnitude than the small anomalous isoscalar magnetic moment of the nucleon $(\mu^{{\rm IS}, (a)}_{N}=-0.06\, \Nmag)$. We recall that we do not consider the singlet component of the magnetic moment operator in Fit 1.
%
%
Comparisons with results of other works are presented in Table \ref{tab:comp}.

We also show the nucleon isoscalar moments associated with the $u$-$d$ and $u$-$d$-$s$ quark currents in Table\ref{tab:strangmag}. 
The corresponding operators are defined by
\begin{eqnarray}
  \hat \mu^{(u+d)}&=&
 \hat \mu_{[1]}+\frac{1}{\sqrt 3}\hat\mu^{8}_{[8]} ,\\
   \hat \mu^{(u+d+s)} &=& \hat\mu_{[1]}.
\end{eqnarray}
The matrix elements of these operators for the nucleon are given in terms of the $F$, $D$, $S$ parameters as
\begin{eqnarray}
   \langle N | \hat\mu^{(u+d)} | N \rangle &=& \left(S + F - \frac{D}{3}\right)
   \Bmag , \\
   \langle N | \hat\mu^{(u+d+s)} | N \rangle &=&  S\, \Bmag .
\end{eqnarray}


\begin{table}
\caption{Nucleon isoscalar moments in units of $\Nmag$. The parenthesis denote the estimated theoretical errors for the last digits. \label{tab:strangmag}}
\begin{ruledtabular}
\begin{tabular}{cccc}
    &  $\mu^{(s)}$ & $\mu^{(u+d)}$& $\mu^{(u+d+s)}$\\
\hline
 Fit 1 & $-0.071$ &$+1.258$ &$+0.815$ \\
 Fit 2  & $+0.155(22)$ & $+1.473(22)$ &$+1.033(22)$\\
 Fit 3  & $+0.161(28)$ & $+1.484(28)$ &$+1.043(28)$\\
\end{tabular}
\end{ruledtabular}
\end{table}

\begin{table*}[htb]
\caption{Comparison of the strange magnetic moment of nucleon between this work, vector meson dominance model~\cite{Jaffe:1989mj}, Skyrme model~\cite{Park:1990is}, Skyrme model with vector mesons~\cite{Park:1991fb}, dispersion relation analysis~\cite{Hammer:1995de,Forkel:1995ff}, meson-exchange model~\cite{Meissner:1997qt}, chiral perturbation theory~\cite{Hemmert:1998pi}, chiral quark soliton model \cite{Kim:1998gt}, QCD equalities~\cite{Leinweber:1995ie} and lattice calculation \cite{Leinweber:2004tc}. 
\label{tab:comp}}
\begin{ruledtabular}
\begin{tabular}{ccccccccccc}
  this work & \cite{Jaffe:1989mj} & \cite{Park:1990is}&
    \cite{Park:1991fb}&\cite{Hammer:1995de} &  
  \cite{Forkel:1995ff}&\cite{Meissner:1997qt}&
      \cite{Hemmert:1998pi} &\cite{Kim:1998gt}& \cite{Leinweber:1995ie}&\cite{Leinweber:2004tc} \\
$+0.155(22)$ & $-(0.25 \div 0.43)$ & $-0.13$ & $-0.05$ &$ -0.24(3) $&
 $-0.26$ & $+0.003$ & $+(0.02\div0.03)$& $+0.41(18)$&$-0.75(30)$ &  
 $-0.046(19)$
\end{tabular}
\end{ruledtabular}
\end{table*}

\section{Discussion}

Several remarks are in order at this point. First, 
the experimental extraction of the proton form factor related to the strange vector current is, in principle, a well-defined procedure. However, the theoretical discussion based on Eq.(\ref{eq:strangeMM}), involves potential ambiguities concerning the way in which SU(3) breaking enters in the magnetic moment operator, as the symmetry breaking effects act differently within the octet and in the mixing with other representations. Irrespective of these principal uncertainties, our conclusion is that the singlet,
decuplets 
and 27-plet anomalous magnetic moments are very small. Consequently, the observed smallness of the isoscalar nucleon magnetic moment implies a strong suppression of the nucleon's strange magnetic moment. This conclusion does not change under different possible SU(3) breaking scenarios.  

In the context of chiral perturbation theory, there are operators in the sector of the singlet current which do not involve the quark mass matrix \cite{Musolf:1996zv}. Since the quark mass terms are the only source of SU(3) flavor breaking in chiral perturbation theory, the singlet moment in the chiral limit contributes negligibly to the octet moments, at higher orders of the electromagnetic coupling, through current matrix elements connecting singlet and  octet. 
One should note, however, that the symmetry breaking scenarios discussed in this paper are in a region quite far removed from the chiral limit, so that singlet-octet mixing potentially involves mechanisms  different from just those driven by the current quark masses.

It is interesting to compare our analysis with other models in which configuration mixing is addressed, such as the Skyrme soliton model.  The present analysis suggests a positive value of the strange magnetic moment of proton, whereas Skyrme models predict negative values \cite{Park:1990is,Park:1991fb}. An alternative recent work \cite{Zou:2005xy} investigates the strange magnetic moment of the  proton in a constituent quark picture, including a configuration $uuds\bar s$, and suggests that the sign of the strange moment reflects the hidden strange quark components in the proton.  The positive value of the strange moment may indicate that the $uuds\bar s$ component preferentially forms a quark cluster rather than a Skyrme type soliton structure with a strong $K^{0}\Lambda$ meson cloud configuration. 

The smallness of the anomalous part of the singlet moment has a correspondence in the SU(6) quark model where the flavor singlet moment is simply represented by the quark operator 
\begin{equation}
  \hat \mu_{q}^{\rm FS} = \sum_{i = 1}^{3} \frac{1}{2m_{i}} {\bf \sigma}_i^z \,\,\,.   
\end{equation}
In the exact SU(3) limit with degenerate quark masses $m_i$, this operator just counts the spin of the baryon. In this limit, $S=1$ and the anomalous flavor singlet moment vanishes. 

The smallness of the anomalous singlet magnetic moment is also consistent with low-energy theorems for the singlet vector current. We first recall the Gerasimov-Drell-Hearn (GDH) sum rule which demonstrates that the $\Delta(1230)$ excitation plays a dominant role in determining the large (isovector) anomalous magnetic moment of the proton. If we consider the corresponding version of the GDH sum rule relevant to the flavor singlet current, then this sum rule suggests strong suppression of the singlet anomalous moment since the flavor singlet current does not couple the baryon octet to the decuplet.

\section{Summary and conclusions}

We have discussed the magnetic moments of the octet baryons in a group theoretical approach without explicit reference to detailed model assumptions about the spin structure of the baryons. We have proposed a separate treatment of the isovector and isoscalar parts of the magnetic moments. Such an 
approach provides model independent relations among the baryon magnetic moments as shown in Eqs.(\ref{eq:relationIV}) and (\ref{eq:relationIS}), based on the octet dominance of the magnetic moment operator. 

For further illustration, we have examined SU(3) breaking scenarios under the assumption that the baryon magnetic moment operator has the same SU(3) group structure as that of the constituents with SU(3) breaking driven by the quark masses.  

From the observation that the anomalous flavor singlet moment is less than one third of the normal part, we conclude that the strange magnetic moment of the proton is small, $\mu^{(s)} \simeq +0.16\, \mu_N$. Its magnitude is less than three times that of the very small anomalous isoscalar magnetic moment of the nucleon. This result is  compatible with the existing experimental data \cite{Aniol:2000at,Aniol:2005zg,Spayde:2003nr,Ito:2003mr,Maas:2004ta,Maas:2004dh}.  
%
Specifically, our small but positive value of $\mu^{(s)} $ is consistent with the recent analysis of Ref. \cite{Aniol:2005zg}.

\begin{acknowledgments}
D.J.\ gratefully acknowledges support of his research in Germany by a Fellowship of the Alexander von Humboldt Foundation. 
\end{acknowledgments}

\appendix
\section{Octet baryon mass in the SU(3) limit}

With the assumption that the SU(3) breaking effects on the baryon mass operator behave as the 8th component of the octet representation, 
the octet baryon masses are parametrized in terms of three quantities, $M_{0}$, $F_{M}$ and $D_{M}$ as
\begin{eqnarray}
  \langle B|H_{M}|B\rangle &= & \Bsymmass\, {\rm Tr}[\bar BB]+  
   F_{M}\, {\rm Tr}\!\left[\bar B [H_{\rm SB},B]\right]\nonumber \\ &&
 +   D_{M}\, {\rm Tr}\!\left[\bar B \{H_{\rm SB},B\}\right]
\end{eqnarray}
where
$
   H_{\rm SB} = \frac{1}{\sqrt3}\lambda^{8}
$
and the baryon matrix $B$ is given in Eq.(\ref{eq:barymat}). The baryon mass in the SU(3) symmetric limit is denoted by $M_{0}$. Each baryon mass is expressed as
\begin{subequations}
\begin{eqnarray}
  M_{N} &=& \Bsymmass + F_{M} - {\textstyle \frac{1}{3}}D_{M}, \\
  M_{\Lambda} &=& \Bsymmass - {\textstyle \frac{2}{3}} D_{M}, \\
  M_{\Sigma} &=& \Bsymmass + {\textstyle \frac{2}{3}} D_{M}, \\
  M_{\Xi} &=& \Bsymmass - F_{M} - {\textstyle \frac{1}{3}} D_{M}.
\end{eqnarray}
\end{subequations}

Fitting the $\Bsymmass$, $F_{M}$, $D_{M}$ parameters from the isospin average masses in the least square method, we obtain
\begin{equation}
\begin{array}{cc}
   \multicolumn{2}{c}{\Bsymmass=1152\ {\rm MeV},}\\
 F_{M} =-190\ {\rm MeV},\ \ \  & D_{M}= 60\ {\rm MeV}.\ \ \
\end{array}
\end{equation}


\end{document}